# Structure of Synchronized Chaos Studied by Symbolic Analysis in Velocity–Curvature Space

A. V. Makarenko

Constructive Cybernetics Research Group, Moscow, Russia

e-mail: avm.science@mail.ru

**Abstract.** A new method of symbolic analysis based on finite discretization of velocity–curvature space is proposed. A minimum alphabet is introduced in a natural way, and a number of initial analytic measures are defined that make it possible to study the structure of discrete mapping dynamics. The proposed method is tested by application to a system of two unidirectionally coupled logistic maps. It is shown that this method can be used to reveal and study changes in the structure of attractors. In the given example, features in the attractor structure of the driven subsystem are studied upon its escape from the identical synchronization regime.

Methods of symbolic dynamics [1–4] offer an effective tool for analysis of dynamical systems determined by discrete maps of the following type:

$$s_{k+1} = f(s_k \ldots s_{k-r}, \mathbf{p}), \quad s \in S \subset \mathbb{R}, \quad k \in K \subseteq \mathbb{Z}, \qquad (1)$$

where $\mathbf{p}$ is the vector of coefficients and $r$ is the memory parameter of the system, $r \geq 0$. Using the Poincare map [5], it is also possible to study continuous systems (flows).

The term "symbolic dynamics" reflects the main idea of this approach, according to which the dynamics of system (1) in space $S$ is described using permissible sequences of symbols (words) $\{C_k\}_{-\infty}^{\infty}$ from a finite or infinite set (alphabet) $\mathfrak{I}$ used to encode the corresponding trajectories $\{s_k\}_{-\infty}^{\infty}$. At present, the symbolic dynamics is used to study various complex phenomena in dynamical systems, including chaos, strange attractors, hyperbolicity, structural stability, controllability of dynamical systems, etc. Applied aspects of symbolic dynamics are also of interest [6, 7].

An original computer-aided method of cell-to-cell mapping for description of dynamical systems in terms of finite alphabets has been proposed by Hsu [8]. A generalized variant of this method leads to finite Markov chains.

Osipenko et al. [9, 10] introduced the concept of a symbolic image of a dynamical system with respect to a finite covering. This image is represented by an oriented graph with the $i$-th vertex corresponding to cell (box) $M_i \in S$ and the $i \to j$ edge present provided that box $M_i$ contains point $s$ such that its image $f(s)$ occurs in box $M_j$.

Dellnitz et al. [11, 12] developed an adaptive subdivision technique for numerical investigation of the temporal behavior of dynamical systems. The idea of this method consists in excluding unnecessary boxes (according to certain conditions) and making more fine subdivision of retained boxes, so that the box refinement leads to more correct approximation.

In order to study the properties of type (1) systems, the alphabet of symbolic dynamics is usually defined on space $S$, which is subdivided into boxes $M_i$ according to certain rules [3, 4, 8, 9, 11]. However, this approach to discretization has the following disadvantages:

  i. The alphabet is not invariant with respect to shear and scaling transformations of the type as $a\,s+b \to s$, which reduces its analytical possibilities for studying systems that are multiplicatively and/or additively nonstationary.

  ii. Determination of symbols by subdividing space $S$ into boxes $M_i$ is not single valued, and, as a rule, each system requires its own scheme of subdivision.

  iii. There are various rules of substituting symbols for trajectories that fall on the box boundary. This uncertainty can lead to different treatments of probabilistic measures defined on a given subdivision of space $S$ into boxes $M_i$.

  iv. The number of symbols in the alphabet is only empirically selected, and there is no formalized criterion for the minimum power of subdivision. This circumstance makes it difficult to evaluate a priori the informative compactness and informativity of a chosen alphabet.

In setting an alphabet of symbolic dynamics, it is possible to pass from a subdivision of space $S$ to subdivision of the velocity–curvature space [13]: $A \times \Phi_0$, $A \ni \alpha^T$, $\Phi_0 \ni \varphi_0^T$. For functions $g(t)$ such that $g \in C^2(T)$, $t \in T \subseteq \mathbb{R}$, the values of velocity $\alpha^T$ and curvature $\varphi_0^T$ are represented by smooth functions [13]; for discrete maps of type (1), the $\alpha^T$ and $\varphi_0^T$ values can be defined via finite differences as follows:

$$\alpha^T = c_{ss}\dot{g}\,,\ \varphi_0^T = \frac{c_{ss}c_{as}\ddot{g}}{1+c_{ss}^2\dot{g}^2}\,,\ \alpha_k^T = c_{ss}(s_{k+1}-s_k)\,,\ \varphi_{0,k}^T = c_{as}\,\mathrm{tg}\frac{\mathrm{arctg}\,\alpha_k^T - \mathrm{arctg}\,\alpha_{k-1}^T}{2}\,, \qquad (2)$$

where $c_{ss}$ and $c_{as}$ are positive coefficients ($c_{ss}, c_{as} > 0$) and the upper dot denotes the derivative with respect to variable $t$.

Proceeding from definition (2), it is possible to introduce in a quite natural way a minimum finite single valued alphabet for $\alpha^T$ and $\varphi_0^T$ on the $A \times \Phi_0$ space, so that it will be free of the aforementioned disadvantages. Indeed, let us define alphabets $C^\alpha$ and $C^\varphi$ for the $\alpha^T$ and $\varphi_0^T$ components, respectively, as follows:

$$C_k^\alpha = \begin{cases} \mathtt{U} & \alpha_k^T > \delta_\alpha^+, \\ \mathtt{D} & \alpha_k^T < \delta_\alpha^-, \\ \mathtt{Z} & \alpha_k^T \in [\delta_\alpha^-, \delta_\alpha^+]. \end{cases}\ ,\ C_k^\varphi = \begin{cases} \mathtt{L} & \varphi_{0,k}^T \in [\delta_\varphi^-, \delta_\varphi^+] \vee C_k^\alpha = C_{k-1}^\alpha = \mathtt{Z}, \\ \mathtt{E} & \varphi_{0,k}^T \notin [\delta_\varphi^-, \delta_\varphi^+] \wedge \{\{C_k^\alpha = \mathtt{D} \wedge C_{k-1}^\alpha = \mathtt{U}\} \vee \{C_k^\alpha = \mathtt{U} \wedge C_{k-1}^\alpha = \mathtt{D}\}\}, \\ \mathtt{B} & \text{otherwise}. \end{cases}$$

Note that, in constructing alphabet $C^\varphi$, we employ the properties of $\varphi_0^T$ as the curvature [13]. The $\delta_\alpha^\mp$ and $\delta_\varphi^\mp$ values are control parameters for the corresponding components, which make it possible to expand the analytical possibilities of the proposed alphabets.

Let us use symbols $C^\alpha$ and $C^\varphi$ to form terms $T^{\alpha\varphi}$ and encode them by $\mathtt{T}\circ$ as follows:

$$T_k^{\alpha\varphi} = \begin{bmatrix} C_k^\varphi \\ C_k^\alpha \end{bmatrix},\ \mathrm{T}^{\alpha\varphi} \ni T_k^{\alpha\varphi},\ \mathrm{T}^{\alpha\varphi}:\ \begin{array}{c|ccc} & \mathtt{B} & \mathtt{L} & \mathtt{E} \\ \hline \mathtt{U} & \mathtt{T5} & \mathtt{T2} & \mathtt{T7} \\ \mathtt{Z} & \mathtt{T4} & \mathtt{T0} & \times \\ \mathtt{D} & \mathtt{T3} & \mathtt{T1} & \mathtt{T6} \end{array}\,, \qquad (3)$$

here $\times$ is an impermissible combination of symbols in the term.

Let us also define an auxiliary alphabet $\tilde{C}^\varphi$ for determining the splitting of terms $\mathtt{T3, T4, T5}$ into subterms $\mathtt{T}\circ\mathtt{P}$ and $\mathtt{T}\circ\mathtt{N}$ as follows:

$$\tilde{C}_k^\varphi = \begin{cases} \text{L} & \varphi_{0,k}^T \in [\delta_\varphi^-, \delta_\varphi^+] \vee C_k^\alpha = C_{k-1}^\alpha = \text{Z}, \\ \text{P} & \varphi_{0,k}^T > \delta_\varphi^+ \wedge \{C_k^\alpha \neq \text{Z} \vee C_{k-1}^\alpha \neq \text{Z}\}, \\ \text{N} & \varphi_{0,k}^T < \delta_\varphi^- \wedge \{C_k^\alpha \neq \text{Z} \vee C_{k-1}^\alpha \neq \text{Z}\}. \end{cases} \quad T^{\alpha\varphi} : \text{T}n \begin{array}{|cc|} \text{P} & \text{N} \\ \hline \text{T}n\text{P} & \text{T}n\text{N} \end{array}, \; n = 3, 4, 5. \quad (4)$$

Now let us form matrix $\mathbf{V}_Q$ of permissible transitions $T_k^{\alpha\varphi} \to T_{k+1}^{\alpha\varphi}$ between terms as shown in Fig. 1a, where unity denotes the allowed transition between terms, a row represents term $T_k^{\alpha\varphi}$, and a column represents term $T_{k+1}^{\alpha\varphi}$. The transition $T_k^{\alpha\varphi} \to T_{k+1}^{\alpha\varphi}$ will be denoted by $Q_k^{\alpha\varphi}$ ($Q^{\alpha\varphi} \ni Q_k^{\alpha\varphi}$) and encoded by symbols $\text{Q}ij$, where $i$ and $j$ are the components of codes $\text{T}i|_k$ and $\text{T}j|_{k+1}$ determined by matrices (3) and (4), respectively. Thus, we obtain graph $\Gamma^{TQ} = \langle T^{\alpha\varphi}, Q^{\alpha\varphi} \rangle$ on the $A \times \Phi_0$ space, representing an extended similitude of a symbolic image of the given dynamical system, which was introduced for space S [9, 10].

In order to quantitatively evaluate characteristics of the structure of sequence $\{s_k\}_{-\infty}^{\infty}$ in $A \times \Phi_0$ space, let us introduce measures $\Delta^\circ$ as the frequencies of appearance of symbols $T^{\alpha\varphi}$ and transitions $Q^{\alpha\varphi}$ in $\{s_k\}_{-\infty}^{\infty}$ ($0 \leqslant \Delta^\circ \leqslant 1$):

$$\Delta^{\alpha\varphi} : \Delta^{\text{T}n} = \frac{|\text{M}^{\alpha\varphi}|}{|\text{K}|}, \quad n = \overline{0,7}, \; \text{M}^{\alpha\varphi} \ni T_k^{\alpha\varphi} : T_k^{\alpha\varphi} = \text{T}n, \quad (5a)$$

$$\Delta^Q : \Delta^{\text{Q}ij} = \frac{|\text{M}^Q|}{|\text{K}|-1}, \quad i,j = \overline{0,7}, \; \text{M}^Q \ni Q_k^{\alpha\varphi} : Q_k^{\alpha\varphi} = \text{Q}ij, \quad (5б)$$

where $|\circ|$ – is the power of the set; the indices $n, i, j$ include subterms (4). Analogous measures can be introduced for symbols $C^\alpha$ and $C^\varphi$.

It should be noted that calculation of the measures introduced by Eqs. (5a) and (5b) makes graph $\Gamma^{TQ}$ labeled, while the derivatives of functions $\Delta^\circ(\delta_\circ^\mp)$ with respect to parameters $\delta_\circ^\mp$ carry an additional important information on the structure of $\{s_k\}_{-\infty}^{\infty}$ in the $S \times K$ space.

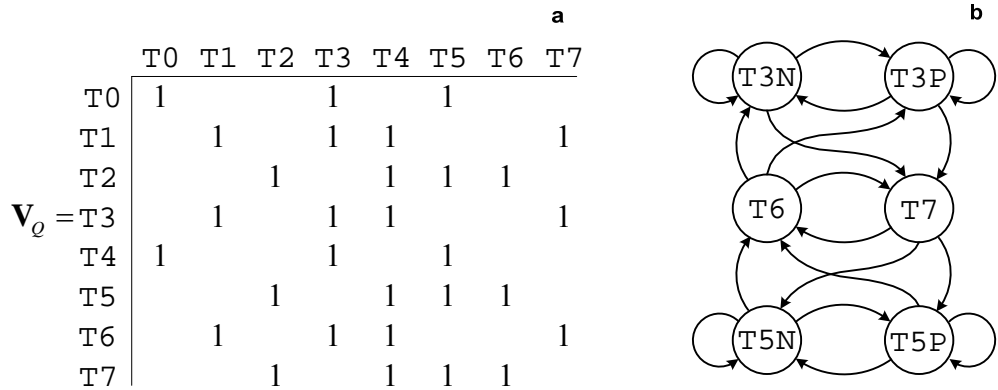

**Fig. 1.** (a) Incidence matrix $\mathbf{V}_Q$ and (b) graph of permissible transitions between terms T6 and T7 terms and T$n$P and T$n$N subterms with $n = 3, 5$ (for $\{\xi_k\}$: $\Delta_\xi^{\text{T}6} = \Delta_\xi^{\text{T}7} \to 1/3$, $\Delta_\xi^{\text{T}n\text{N}} = \Delta_\xi^{\text{T}n\text{P}} \to 1/12$, $\Delta_\xi^{67} = \Delta_\xi^{76} \to 5/24$, $\Delta_\xi^{63\text{N}} = \Delta_\xi^{63\text{P}} = \Delta_\xi^{75\text{N}} = \Delta_\xi^{75\text{P}} \to 1/16$, $\Delta_\xi^{n\text{N}n\text{N}} = \Delta_\xi^{n\text{P}n\text{P}} \to 1/144$, $\Delta_\xi^{n\text{N}n\text{P}} = \Delta_\xi^{n\text{P}n\text{N}} \to 1/72$).

The proposed approach to symbolic dynamics on the $A \times \Phi_0$ space was applied to analysis of the structure of oscillations in a system of two unidirectionally coupled logistic maps [14, 15] during their escape from the identical synchronization regime:

$$x_{k+1} = 4\lambda x_k(1-x_k), \quad y_{k+1} = 4\lambda\left[y_k + \gamma(x_k - y_k)\right]\left(1 - \left[y_k + \gamma(x_k - y_k)\right]\right), \quad (6)$$

where $x_k$ and $y_k$ are the variables of state of the driving and driven process, respectively, $x, y \in [0, 1]$; $\gamma$ is the coupling parameter, $\gamma \in [0, 1]$; and $\lambda$ is the control parameter, that determines the regime of oscillations, $\lambda \in [0, 1]$. The logistic mapping is well known [16] and used as a test model object in nonlinear and chaotic dynamics [5].

The behavior of system (6) was analyzed for $\delta_\alpha^\circ = \delta_\varphi^\circ = 0$. The estimations of $\Delta^\circ$ values were calculated on the interval of $k \in \overline{1 \times 10^5, 4 \times 10^5}$. This shift from $k = 0$ is related to the necessity of minimizing the parasitic influence of a transient process. In addition, all estimations of the analyzed values were averaged over 300 variants of initial conditions $x_0 = \xi_1$, $y_0 = \xi_2$, where $\xi_1, \xi_2 \in (0, 1)$ are uncorrelated uniformly distributed random values. This averaging ensured neutralization of the memory effect induced by the initial conditions on the trajectories of $\{x_k\}$ and $\{y_k\}$ processes. The coupling parameter was varied in the interval of $\gamma \in [0, 0.5]$ at a discretization step of $1 \times 10^{-4}$. The control parameter was set at $\lambda = 0.95$, which corresponded to a regime of developed chaos in system (6) [5]. This choice was explained by the wish to make possible a mutual analysis and ensure the consistency of results obtained in this study and published data [14, 15]. The boundary of symbol detection, which is related to the volume of computations, corresponded to a level of $\breve{\Delta}^\circ = 1.(1) \times 10^{-8}$.

According to published data [14, 15], the escape of system (6) from the identical synchronization regime begins at $\gamma \leq 0.35$, while $\gamma = \gamma_2 = 0.14$ corresponds to a rearrangement of the attractor. At $\gamma = \gamma_{33} \approx 0.2606$ (see [15, Fig. 2]), the structure of attractor for $\{y_k\}$ is also significantly different from that for the free system ($\gamma = 0$).

Analysis showed that the trajectory of the driven system consists predominantly of peak-shaped pulses (terms T6 and T7) with a small admixture of terms T3 and T5 (Fig. 2), which is consistent with our conclusions drawn previously [15]. Note that terms T3 and subterm T5P are only encountered on the interval of $\gamma \in (0, 0.35)$. Thus, the $\{y_k\}$ sequence is significantly asymmetric relative to terms T3 and T5 and their subterms (Fig. 2). The value of $\gamma = \gamma_{33}$ corresponds to an almost maximum concentration of subterm T3N ($\Delta^{T3N} \approx 0.03$) in the $\{y_k\}$ sequence.

As can be seen from Fig. 3a, transition T3 $\to$ T3 is a rather strict structural invariant that is retained in the presence of external noise with intensity on the order of $10^{-3}$. The fraction of this transition in the $\{y_k\}$ sequence is maximum at $\gamma = \gamma_{33}$ and it completely vanishes at $\gamma \geq \hat{\gamma}_{33} \approx 0.2755$. The change in $\Delta^\varrho | \eta$ for transition T5 $\to$ T5 in the region of $\gamma > \gamma_t \approx 0.3365$ is related to nonrobustness of the synchronization regime [14]. Figure 3b reveals a significant asymmetry of the transitions between subterms T3P and T3N, the main contribution to $\Delta^{33}$ at $\gamma = \gamma_{33}$ being due to transition T3P $\to$ T3N. For the comparison, the legend to Fig. 1 presents $\Delta^\circ$ values for $\{\xi_k\}$ – an uncorrelated random sequence with uniform distribution.

Thus, Figs. 2 and 3 clearly demonstrate degeneracy of the degree of stochasticity [17] for oscillations of the logistic map and reveal peculiarity of the structure of a trajectory at $\lambda = 0.95$ not only for the driven subsystem $\{y_k\}$, but for the free oscillator ($\gamma = 0$) as well. This circumstance is manifested by a "poor" ratio of terms T3 and T5 and their subterms with respect to both $\Delta^{\alpha\varphi}$ and $\Delta^\varrho$ relative to those for the $\{\xi_k\}$ sequence.

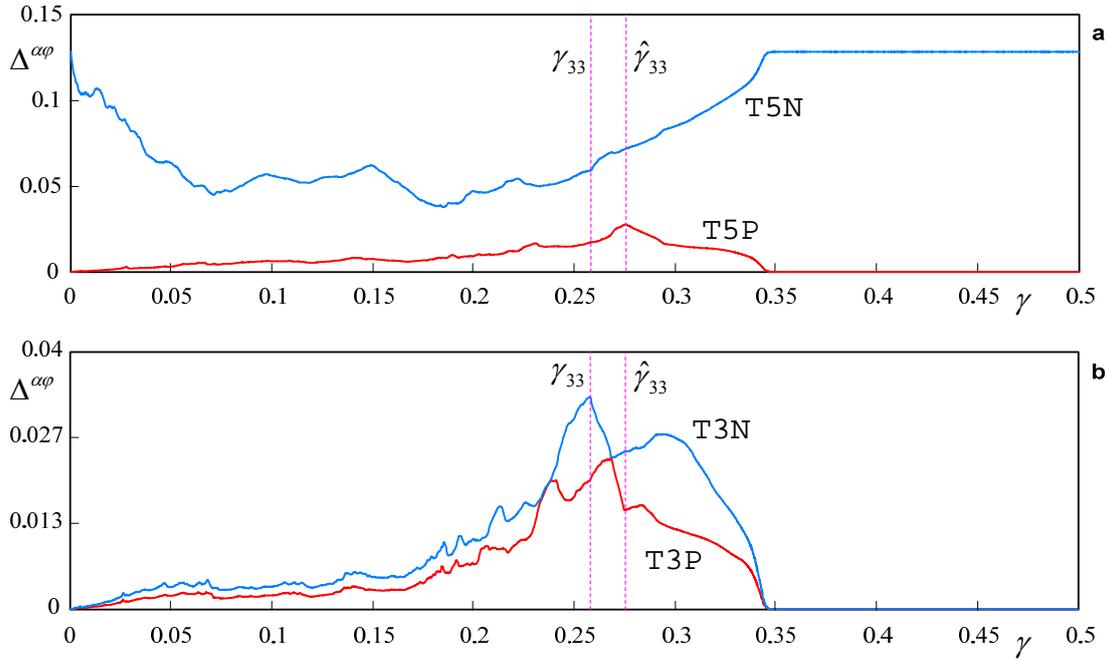

**Fig. 2.** Plots of $\Delta^{\alpha\varphi}$ vs. coupling parameter $\gamma$ for terms (a) T5 and (b) T3.

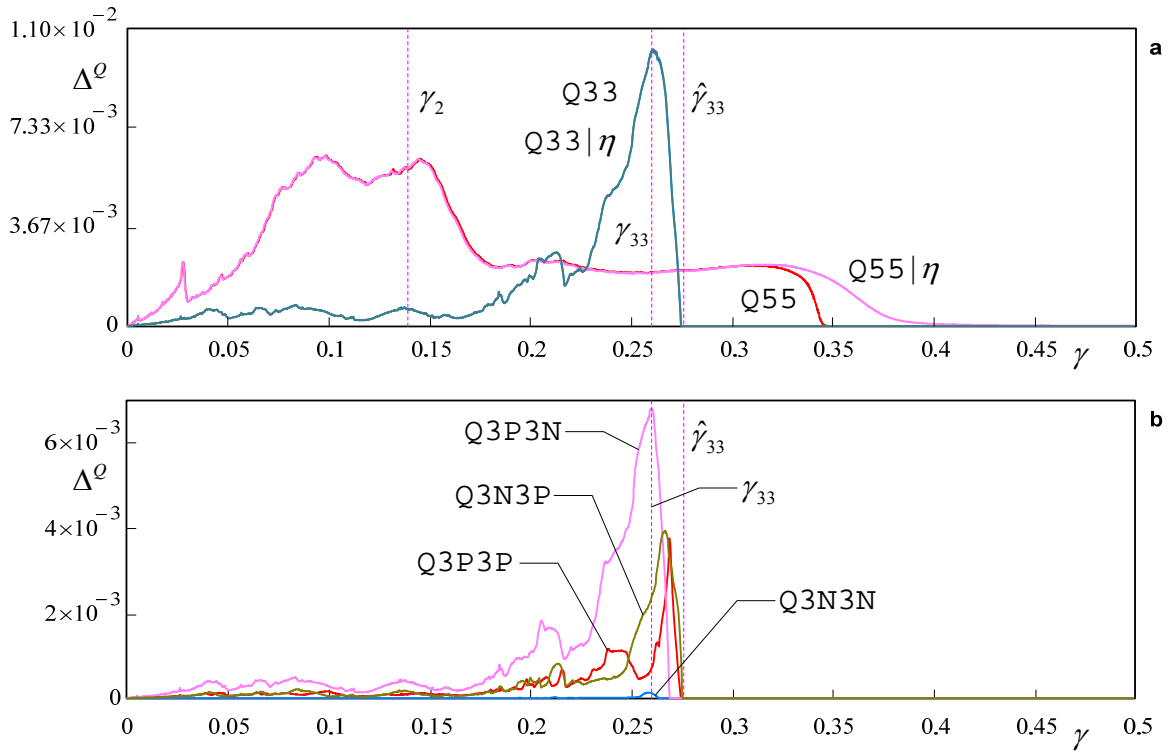

**Fig. 3.** Plots of $\Delta^Q$ vs. coupling parameter $\gamma$ for transitions (a) T3 → T3, T5 → T5 and (b) T3$n$ → T3$m$, $n, m = $ N, P; estimations for Q33|$\eta$ and Q55|$\eta$ upon the additive introduction of a small noise into the driven subsystem: $x_k \to x_k + \eta\left(\xi'_k - x_k\right)$, $\eta = 10^{-3}$, $\xi'_k \in [0, 1]$ (uncorrelated uniformly distributed random values).

Thus, we have proposed a new description of symbolic dynamics based on subdivision of the velocity–curvature ($A \times \Phi_0$) space and introduced a minimum alphabet on this space. The proposed approach (i) is free of disadvantages inherent in the symbolic analysis with subdivision of space S and (ii) makes it possible to study in detail the shape (structure of geometry) of the trajectories of discrete maps in the S × K space (for the importance of this characteristic, see [17] and references therein).

The obtained results demonstrate informativity of the symbolic analysis in the $A \times \Phi_0$ space as applied to investigations of the structure of oscillations in synchronizing dynamical systems. A significant asymmetry in the structure of oscillations, which leads to degeneracy of the degree of their stochasticity, has been found for a model system under consideration. Using the proposed analytical tool, it is possible to reveal details in the structure of an attractor of a driven dynamical subsystem and study the character of external action on this subsystem.

It is planned to expand the analytical possibilities of the proposed technique and apply it to studying multidimensional systems and solving problems of controlling chaotic dynamics and suppressing chaotic oscillations by means of weak external actions.

*Translated by P. Pozdeev*

**Andrey V. Makarenko** – was born in 1977, since 2002 – Ph. D. of Cybernetics. Founder and leader Research & Development group "Constructive Cybernetics". Author and coauthor of more than 50 scientific articles and reports. Associate Member IEEE (IEEE Systems, Man, and Cybernetics Society Membership). Research interests: analysis of the structure dynamic processes, predictability; detection, classification and diagnosis is not fully observed objects (patterns); synchronization in nonlinear and chaotic systems; system analysis and modeling of economic, financial, social and bio-physical systems and processes; system approach to development, testing and diagnostics of complex information-management systems.